%% file: eprint.tex
\documentclass[10pt]{article}
\usepackage{graphicx}
\def\Title#1{\begin{center} {\Large #1 } \end{center}}
\def\Author#1{\begin{center}{ \sc #1} \end{center}}
\def\Address#1{\begin{center}{ \it #1} \end{center}}

\newcommand\pubblock{\rightline{\begin{tabular}{l} Proceedings of the Fifth Annual LHCP\\ \pubnumber\\
         \pubdate  \end{tabular}}}

\newenvironment{Abstract}{\begin{quotation} \begin{center} 
             \large ABSTRACT \end{center}\bigskip 
      \begin{center}\begin{large}}{\end{large}\end{center} \end{quotation}}

\newenvironment{Presented}{\begin{quotation} \begin{center} 
             PRESENTED AT\end{center}\bigskip 
      \begin{center}\begin{large}}{\end{large}\end{center} \end{quotation}}

\def\Acknowledgements{\bigskip  \bigskip \begin{center} \begin{large}
             \bf ACKNOWLEDGEMENTS \end{large}\end{center}}
\input econfmacros.tex
\usepackage{color}

\newcommand\pubnumber{ ATL-PHYS-PROC-2017-101 }
\newcommand\pubdate{\today}

\def\affiliation{
On behalf of the ATLAS and CMS collaborations, \\
Department of Physics and Astronomy \\
University of Sussex, Brighton, BN1 9RH, United Kingdom}

\begin{document}

% large size for the first page
\large
\begin{titlepage}
\pubblock

%% Change the title, name, abstract
%% Title 
\vfill
\Title{  Top quark mass, spin and decay properties }
\vfill

%  if you need to add the support use this, fill the \support definition above. 
%   \Author{ FIRSTNAME LASTNAME \support }
\Author{ Kerim Suruliz  }
\Address{\affiliation}
\vfill
\begin{Abstract}
I review recent results from the ATLAS and CMS collaborations on the measurements of the top quark mass, its spin and decay properties. 
% The fifth annual Large Hadron Collider Physics (LHCP) conference will be held at Shanghai Jiao Tong University in Shanghai from May 15-20, 2017. 

\end{Abstract}
\vfill

% DO NOT CHANGE 
\begin{Presented}
The Fifth Annual Conference\\
 on Large Hadron Collider Physics \\
Shanghai Jiao Tong University, Shanghai, China\\ 
May 15-20, 2017
\end{Presented}
\vfill
\end{titlepage}
\def\thefootnote{\fnsymbol{footnote}}
\setcounter{footnote}{0}
%

% normal size for the rest
\normalsize 

%% Your paper should be entered below. 

\section{Introduction}

Measurements of the properties of the top quark are a crucial part of the LHC physics programme. Top quark pairs are produced at the 13 TeV LHC with a cross section of around 0.8 nb, implying that almost 30 million top quark pair events were produced thus far in the 13 TeV run of the LHC.

Such a large data sample presents an excellent opportunity for precision studies of the properties of the top quark. These studies are well motivated by a number of theoretical arguments. The top quark mass is a key parameter in global fits of electroweak observables, which are an important self-consistency test of the Standard Model. Knowing the top quark mass precisely in turn makes these self-consistency tests more stringent. The coupling between the top quark and the Higgs boson is a leading source of quantum corrections to the Higgs potential, which determines the (meta-)stability of the Universe we live in. The top quark has a width larger than the characteristic energy scale of quantum chromodynamics (QCD), which means that it decays before it hadronises. Therefore, unlike the case of other quark flavours, there is no bound state of top quarks (``toponium''), and studying the top quark gives us insight into the behaviour of a bare quark. 

In this article I will review recent measurements from the ATLAS~\cite{Aad:2008zzm} and CMS~\cite{Chatrchyan:2008aa} collaborations concerning the spin, decay properties and mass of the top quark. 

%$\sigma_{t\bar{t}}=832^{+40}_{-46}$ pb 

\section{Top quark spin}
Since the initial state quarks and gluons are unpolarised in $pp\to t\bar{t}$ production, top quarks are also produced unpolarised at the LHC~\footnote{Apart from $<1\%$ contributions to the polarisations arising from absorptive parts of the scattering amplitudes.}. However, QCD predicts that there is a small correlation between the spins of the top and the anti-top in $t\bar{t}$ events. Measurements of spin correlation in $t\bar{t}$ events by the ATLAS and CMS collaborations (see e.g. \cite{Aad:2014mfk,Khachatryan:2016xws}) provide direct evidence that the top quark does not hadronise before decaying. 

In a recent measurement~\cite{Aaboud:2016bit}, ATLAS went one step further and measured the spin part of the amplitude for $t\bar{t}$ production at the LHC. The normalised double-differential cross section for $t\bar{t}$ production at the LHC can be written as:
\begin{equation}
\frac{1}{\sigma}\frac{d^2\sigma}{d\cos{\theta^a_+}d\cos{\theta^b_-}} = \frac{1}{4} (1+B^a_+\cos{\theta^a_+} + B^b_-\cos{\theta^b_-} - C(a,b)\cos{\theta^a_+}\cos{\theta^b_-}).
\end{equation}
Here $\theta^a_{\pm}\ (a=1,2,3)$ are angles between decay products (taken to be leptons in this measurement) of the top and anti-top and suitably chosen axes, in the rest frame of the top or anti-top. $B^a_{\pm}$ are polarisation coefficients, and are predicted to be close to zero, while $C(a,b)$ are spin correlation coefficients. Previously, a single spin correlation coefficient along one axis - the so called {\it helicity axis} - was measured.

\begin{figure}[htb]
\centering
\includegraphics[height=2in]{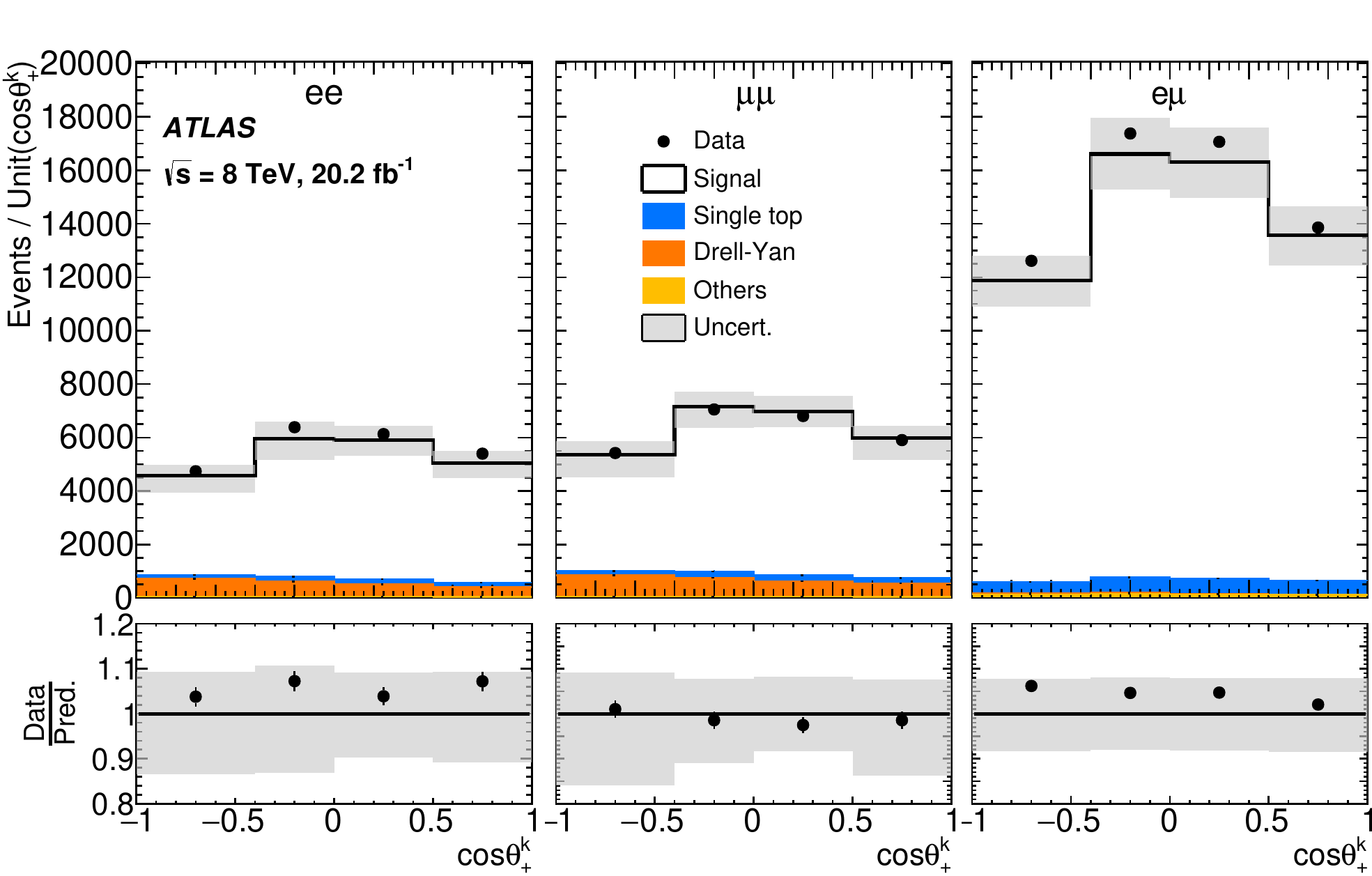}
\includegraphics[height=2in]{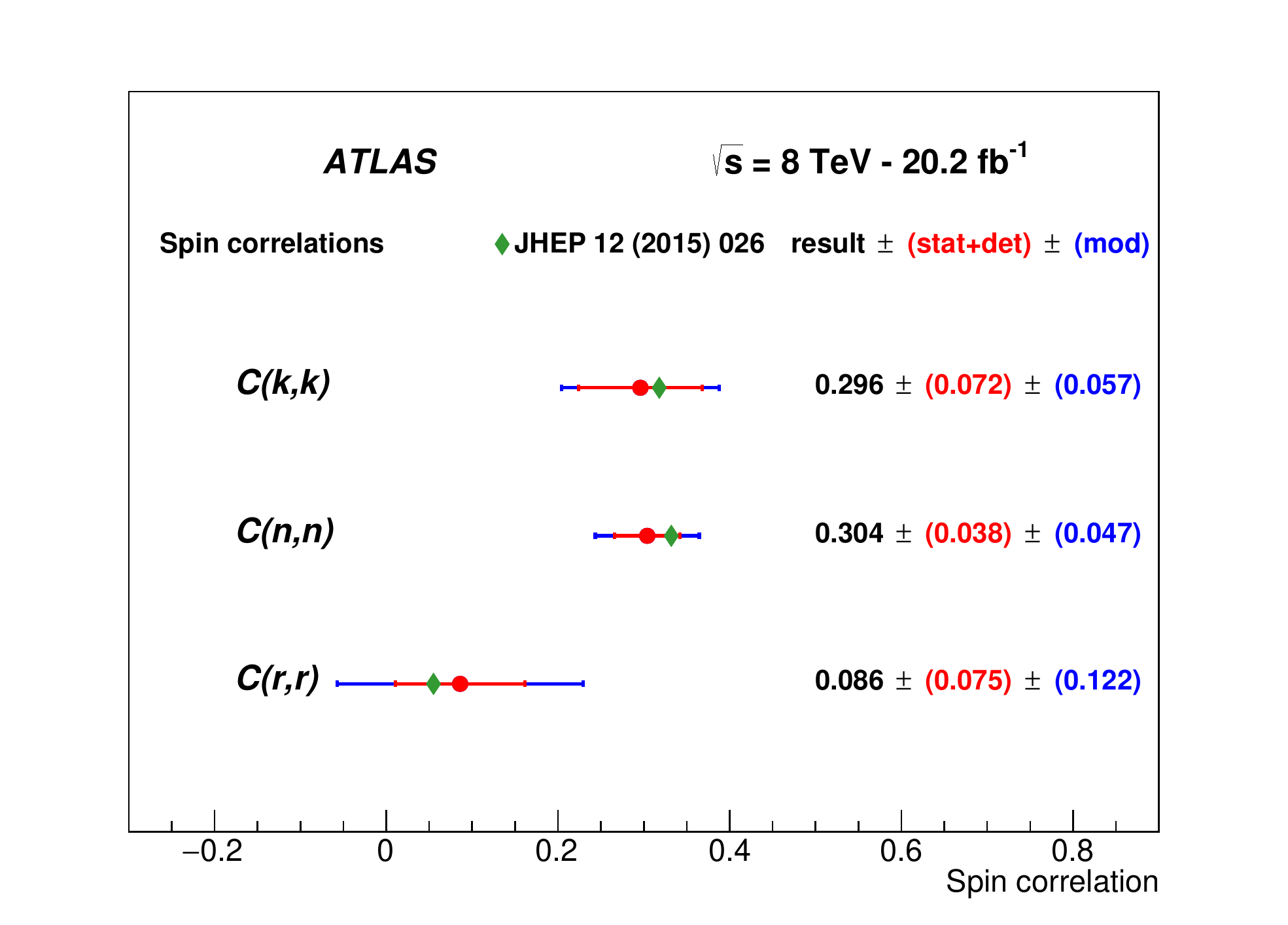}
\caption{(left) Distributions of $\cos\theta^+$ with respect to the helicity axis ($k$) in data and simulation. (right) The measured values of (a subset of) the spin correlations for the parton-level measurement, compared to the predictions from the Standard Model. The figures are taken from Ref.~\cite{Aaboud:2016bit}.}
\label{fig:ATLASspinobservables}
\end{figure}

The measurement, whose objective is to extract all 15 coefficients $B^a_{\pm}$ and  $C(a,b)$, is performed in the dilepton channel using 8 TeV data collected by ATLAS. In order to measure the angles $\theta^a_{\pm} (a=1,2,3)$, the $t\bar{t}$ system is reconstructed using the neutrino weighting technique introduced in Ref.~\cite{Abbott:1997fv}. A Bayesian method is used to unfold the reconstructed distributions to the parton and particle level where results may be directly compared with the Standard Model prediction. An example of the reconstructed distributions and extracted spin correlation coefficients is shown in Figure~\ref{fig:ATLASspinobservables}. No significant deviations from the theoretical expectations are found. As is the the case for most measurements reviewed here, the signal modelling uncertainties typically form the leading contribution to the total uncertainty in the measurement.

\section{Top quark decays}
The standard model predicts that the top quark decays almost exclusively to a $W$ boson and the $b$-quark, as dictated by the structure of the CKM matrix. The $Wtb$ vertex is of V-A type in the Standard Model; its structure sets the relative fractions of polarisations of $W$ bosons arising from top quark decays. Any observed deviation from these fractions would indicate that the $Wtb$ vertex is modified by contributions from physics beyond the Standard Model (BSM). 

In the $W$ boson rest frame, the normalised differential cross section of an analyser is given by:
\begin{equation}
\frac{1}{\sigma}\frac{d\sigma}{d\cos{\theta^*}} = 
\frac{3}{4} (1-\cos^2{\theta^*}) F_0 + \frac{3}{8}  (1-\cos {\theta^*})^2 F_L + \frac{3}{8}  (1+\cos {\theta^*})^2 F_R,
\end{equation}
where $\theta^*$ is the angle between the analyser and the reversed direction of flight of the $b$-quark from the top quark decay in the $W$ boson rest frame. $F_0, F_L$ and $F_R$ are the polarisation fractions.

ATLAS presented a measurement of the $W$ polarisation fractions in Ref.~\cite{Aaboud:2016hsq} using the full 8 TeV dataset. The measurement is performed in the $l$+jets channel. The $t\bar{t}$ system is reconstructed using a kinematic likelihood method, and two different analysers are explored: the charged lepton from the leptonically decaying $W$ boson and the $d$-quark of the hadronically decaying $W$ boson. Due to the difficulties in identifying the $d$-type quark with sufficiently high efficiency and the larger sensitivity of the hadronic analyser measurement to modelling uncertainties, the measurement using the leptonic analyser is found to have significantly better sensitivity.

\begin{figure}[htb]
\centering
\includegraphics[height=2in]{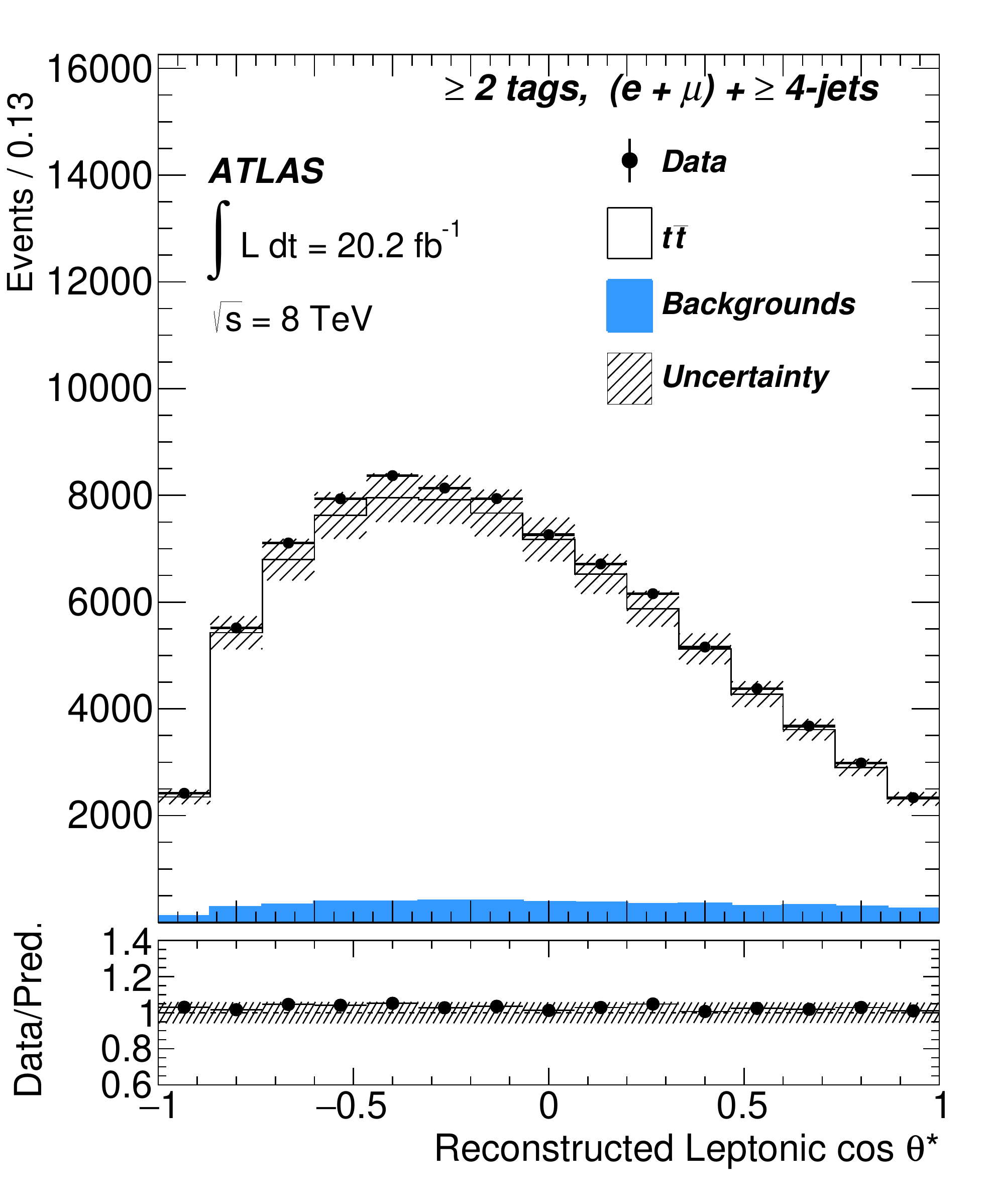}
\includegraphics[height=2in]{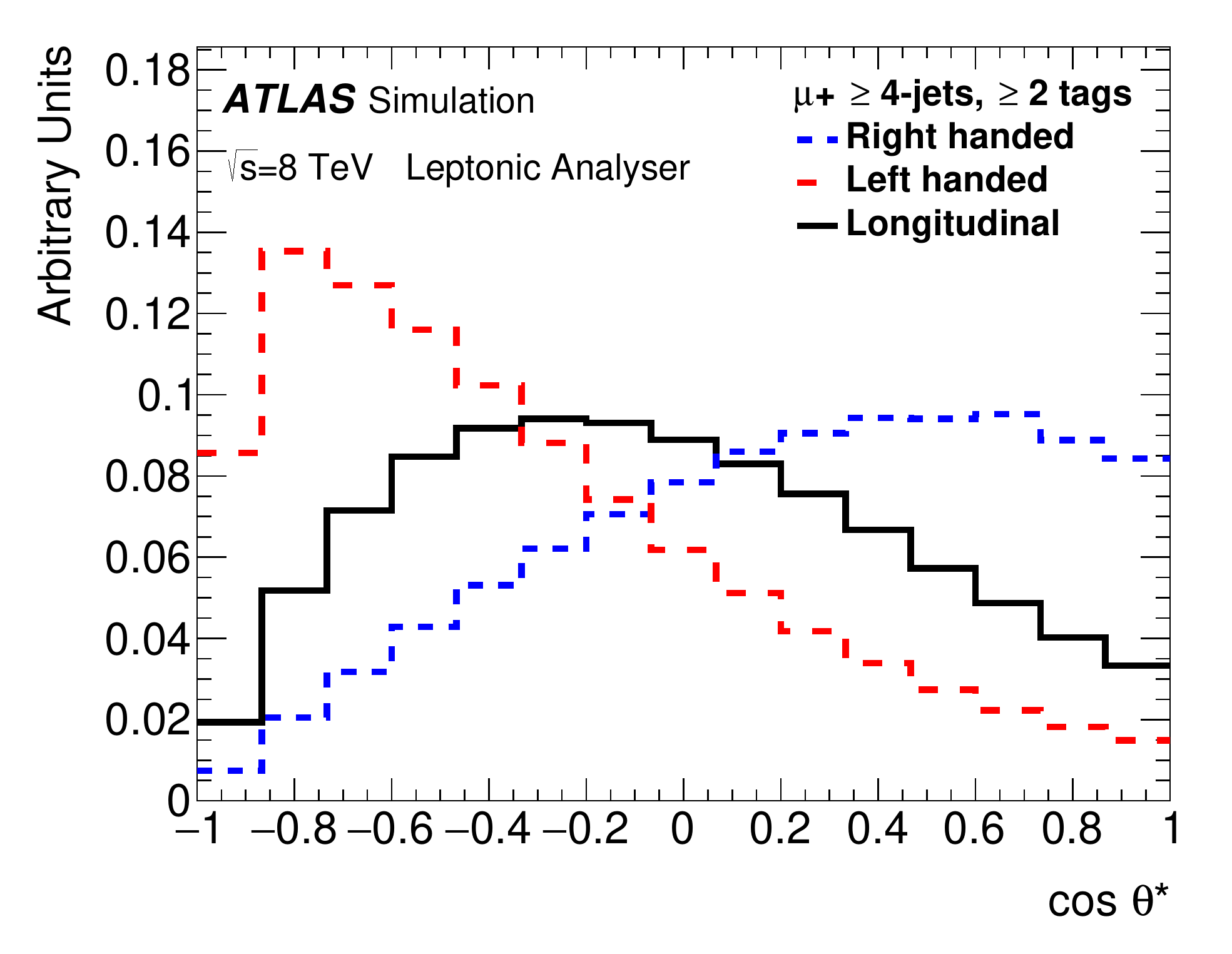}
\caption{(left) The reconstructed $\cos{\theta^*}$ distribution in data compared to the expectation from simulated signal and background. (right) Templates of the $\cos{\theta^*}$ distribution for the different polarisation fractions, for the leptonic analyser. The figures are taken from Ref.~\cite{Aaboud:2016hsq}.}
\label{fig:ATLASWhelicity}
\end{figure}

The $W$ polarisation fractions are extracted from the measured $\cos{\theta^*}$ distribution using a template method. The good agreement between the measured distribution and the expectations are shown in Figure~\ref{fig:ATLASWhelicity}. The figure also shows the templates for the different polarisation fractions for the leptonic analyser. The measured fractions are found to be in agreement with the Standard Model predictions; limits are set on anomalous couplings that parametrise generic modifications of the $Wtb$ vertex. The measurement is dominated by systematic uncertainties, in particular the jet energy scale and resolution. CMS also measured the $W$ polarisation fractions in the lepton+jets channel with the 8 TeV dataset~\cite{Khachatryan:2016fky}, with similar precison.

The Standard Model predicts negligible CP violation in the top quark sector. However, baryogenesis requires sources of CP violation beyond the SM, and many extensions of the SM also predict new CP violating phases in the top quark sector. CMS has sought for signals of such new sources of CP violation, using $t\bar{t}$ events in the lepton+jets channel in the full 8 TeV dataset~\cite{Khachatryan:2016ngh}.

This measurement uses observables of the form $O_i\equiv \vec{v}_1\cdot (\vec{v}_2\times \vec{v}_3)$, where $\vec{v}_i$ are 3-vectors formed from reconstructed final state particle momenta, in the appropriate frame. Such triple products are CP-odd and therefore sensitive to possible CP violation effects. The asymmetry 
\begin{equation}
A_{CP} (O_i) = \frac{N_{\rm events} (O_i>0) - N_{\rm events} (O_i<0)}{N_{\rm events} (O_i>0) + N_{\rm events} (O_i<0)}
\end{equation}
is measured, and a significant difference from zero in the obtained result would indicate the presence of CP violation. 

Four different observables are considered, constructed from the momenta of the three-momenta of the $b,\bar{b},l$, and non-$b$ quark jet ($j_1$) originating from the hadronically decaying $W$ boson with the highest transverse momentum. A $\chi^2$ procedure is used to reconstruct the $t\bar{t}$ system, and the charge of the lepton is used to differentiate between the $b$ and $\bar{b}$ jets.

The signal and background yields are obtained from a fit to the $M_{lb}$ distribution, where $b$ denotes the $b$-jet from the semileptonically decaying top quark. The shape of the background distribution is obtained from a control region which is similar to the signal region, but requiring no $b$-tagged jets. The asymmetry in the background, which is dominated by $W$+jets events, is found to be consistent with zero in the control region. 

\begin{figure}[htb]
\centering
\includegraphics[height=2in]{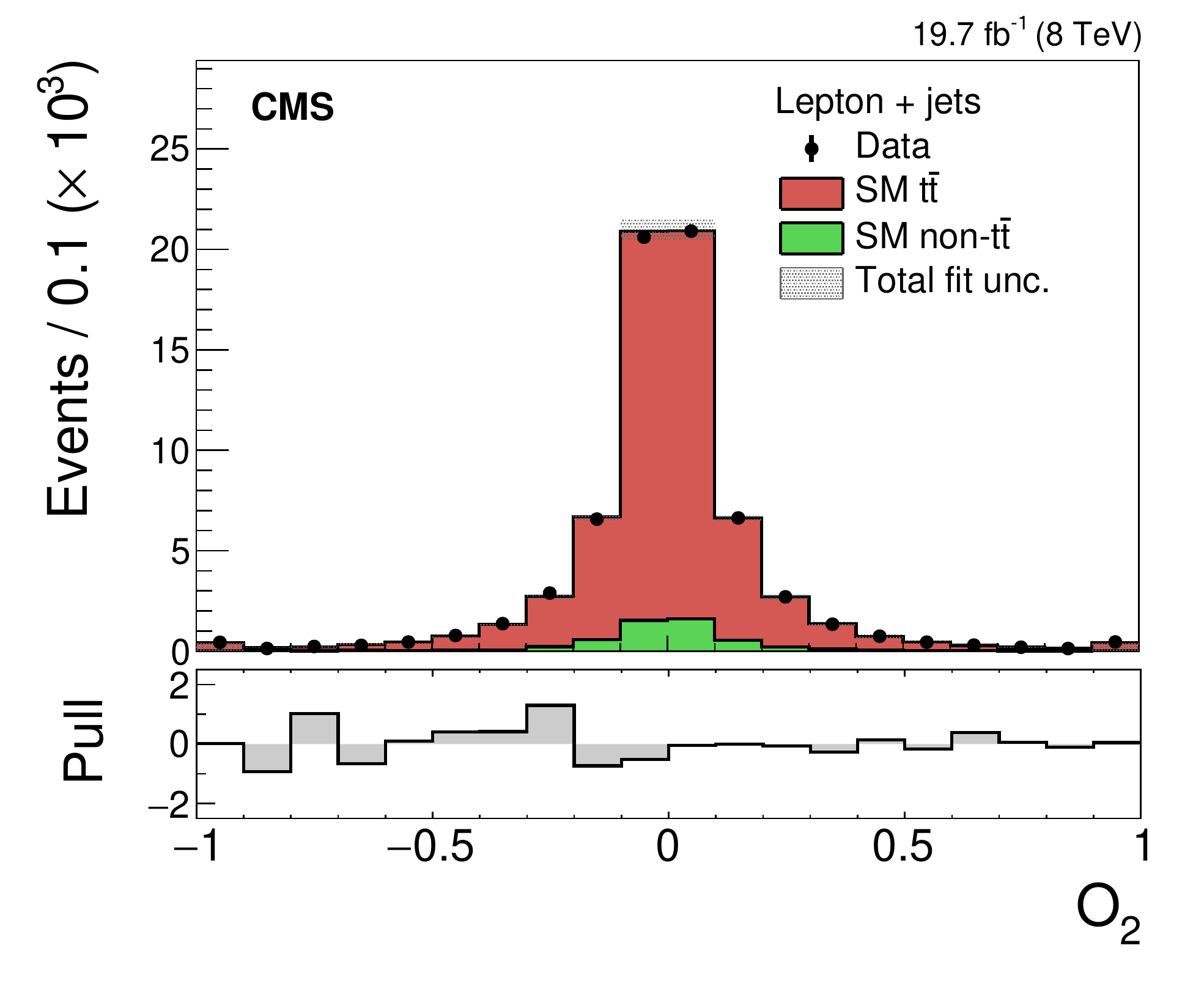}
\includegraphics[height=2in]{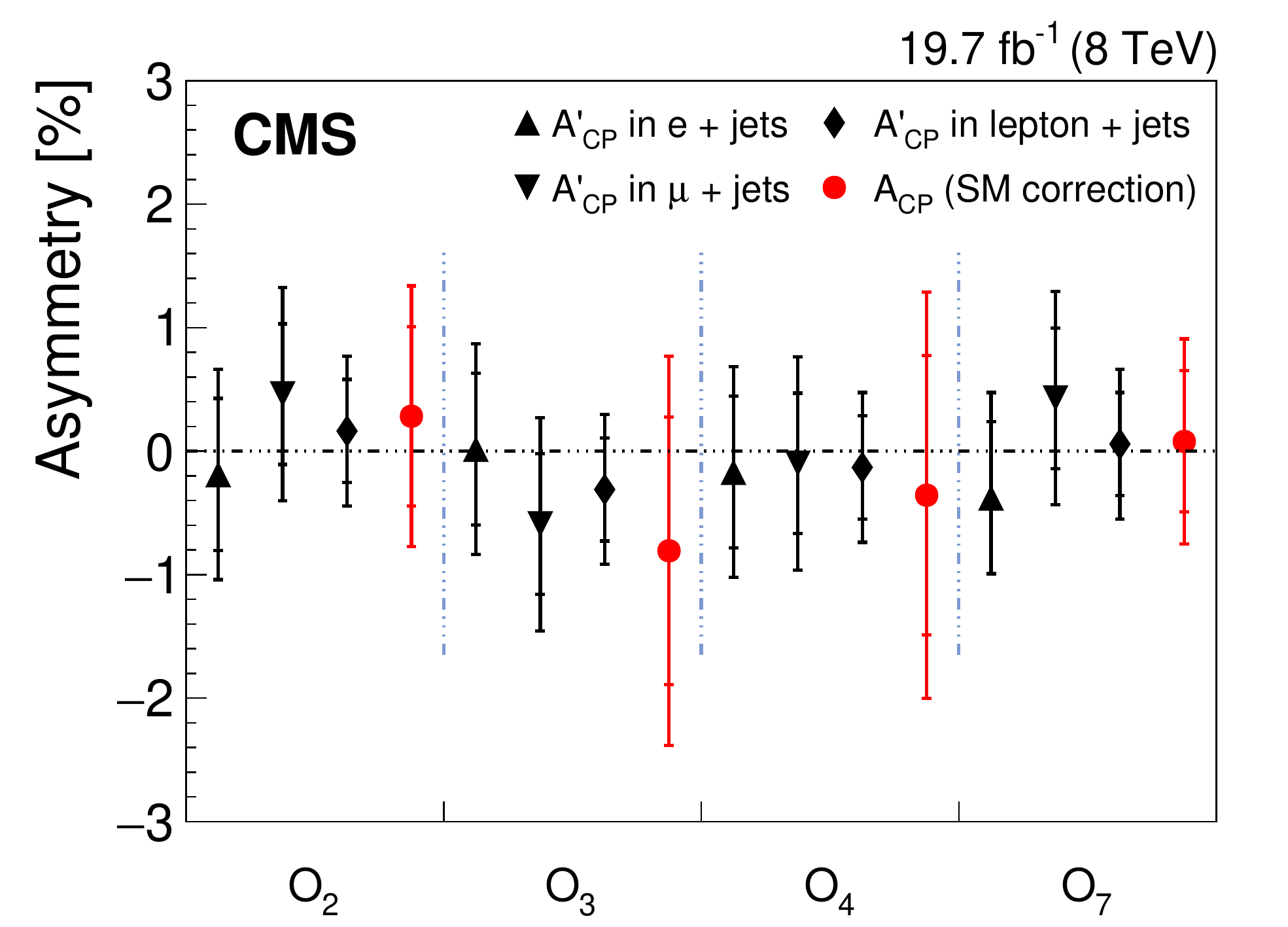}
\caption{(left) Distribution of one of the CPV observables considered, $O_2\equiv (\vec{p}_b+\vec{p}_{\bar{b}})\cdot (\vec{p}_l\times\vec{p}_{j_1})$, compared to simulated signal and background normalised to event yields obtained by performing the fit to $M_{lb}$. (right) The measured asymmetries for all observables considered. The figures are taken from Ref.~\cite{Khachatryan:2016ngh}.}
\label{fig:CMSCPviolation}
\end{figure}

The asymmetries for all the observables considered are found to be consistent with zero, as shown in Figure~\ref{fig:CMSCPviolation}. The dominant uncertainty arises from the limited statistics in the control region, which is used to estimate the potential bias in the asymmetry arising from the background.

Another measurement of CP violation in $t\bar{t}$ events was presented by the ATLAS collaboration in Ref.~\cite{Aaboud:2016bmk}. Rather than searching for CP violation in the top quark sector itself, this measurement uses the fact that $b$ quarks and hence $b$-hadrons are copiously produced in top quark decays via $t\to Wb$, and that the charge of the $b$-hadrons at production can be determined by reconstructing the $t\bar{t}$ system. This, combined with the use of a {\it soft muon tagger} (SMT) to determine the charge of the $b$-hadron at decay, renders possible the measurement of CP violation in $b$-hadrons using $t\bar{t}$ events.

The measurement is performed in the $l$+jets channel and uses a kinematic likelihood to reconstruct the $t\bar{t}$ system. Similarly to the CMS measurement of CP violation described earlier, the charge of the lepton is used to differentiate between the $b$ and $\bar{b}$ jets. At least one of the $b$-jets is required to be tagged by the soft muon tagger, which requires the presence of a muon close to the jet. The charge of the muon is used to determine the charge of the $b$-hadron at decay. Asymmetry observables are constructed based on the number of observed events with either charge of the SMT muon and the lepton from $W$ boson decay.

The measured asymmetries are found to be compatible with zero. The dominant uncertainties are statistical, and this measurement will benefit significantly from the larger dataset available in the $\sqrt{s}=13$ TeV run of the LHC. To compare with theoretical expectations, the results are unfolded to particle level in a fiducial region. Limits are set on the sizes of CP violation parameters, in some cases improving on the current state of the art limits coming from B-factories. 

\section{Top quark mass}
The top quark mass is measured in a variety of ways at the LHC experiments. Broadly speaking, the measurements can be classified
into one of three categories: (i) direct measurements in $t\bar{t}$ events; (ii) alternative methods (such as using single top quark events or events with top quarks decaying to $b$-jets containing $J/\psi$'s) and (iii) measurements of the pole mass of the top quark using the $t\bar{t}$ cross section or $t\bar{t}$+1 jet events. 

Direct measurements of the top quark mass in $t\bar{t}$ events have been carried out in the dilepton, $l$+jets and all-hadronic channels. They typically rely on the template method, with templates obtained from Monte Carlo simulation, to extract the top quark mass. This raises questions about the relationship of the measured quantitity to the physical parameter of interest, the top quark pole mass\footnote{Interesting recent developments in addressing this issue were presented by J. Lindert and M. Preisser at this conference.}. 

%We now turn to discussing recent ATLAS and CMS results on the top quark mass. 

%The most precise quark mass measurements are dominated by systematic uncertainties.

In Ref.~\cite{Aaboud:2017mae} ATLAS reported a measurement of the top quark mass in the all-hadronic channel. This is a challenging measurement, with a large combinatorial background as well as a background from QCD multijet production. 

\begin{figure}[htb]
\centering
\includegraphics[height=2in]{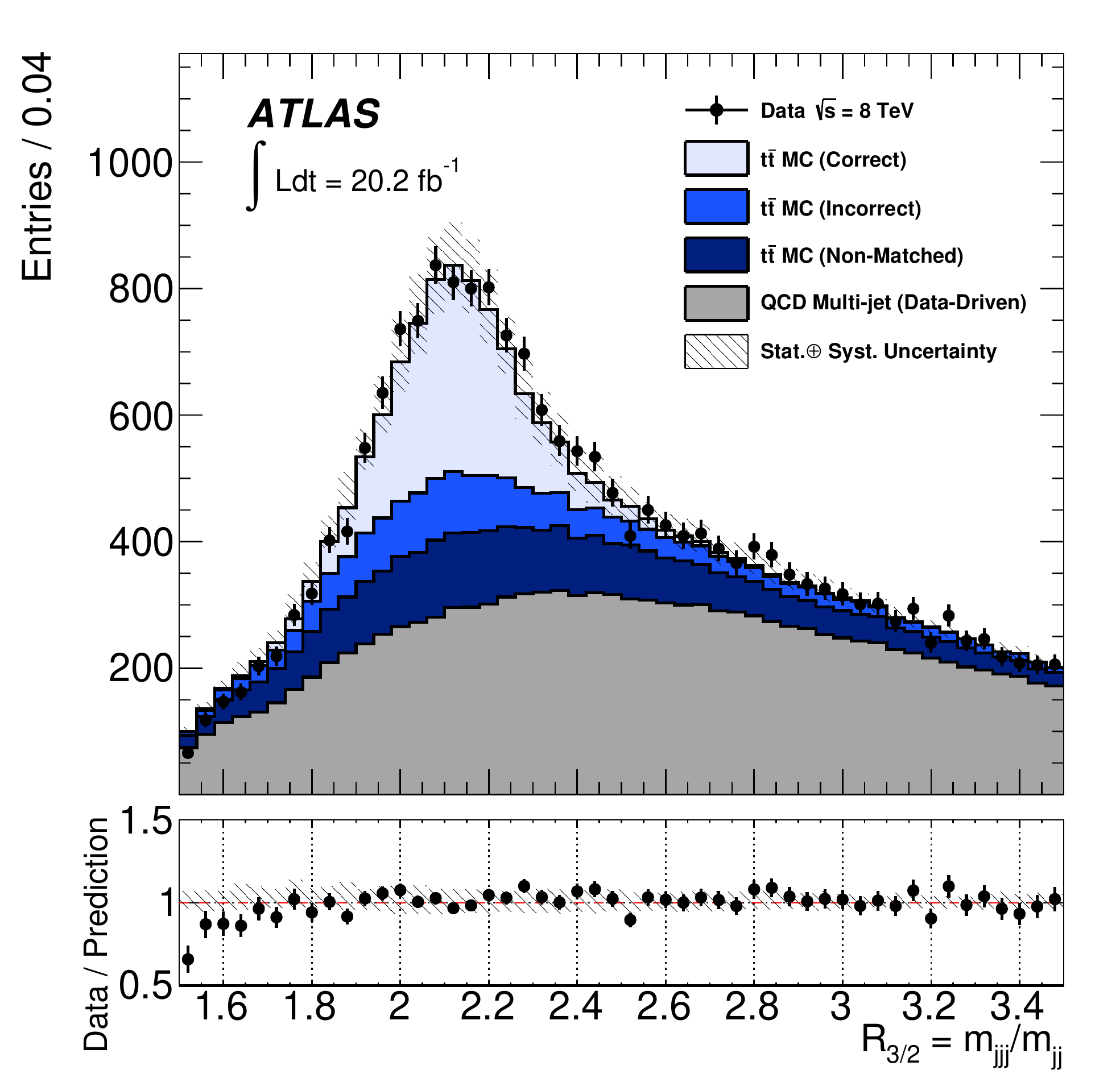}
\includegraphics[height=2in]{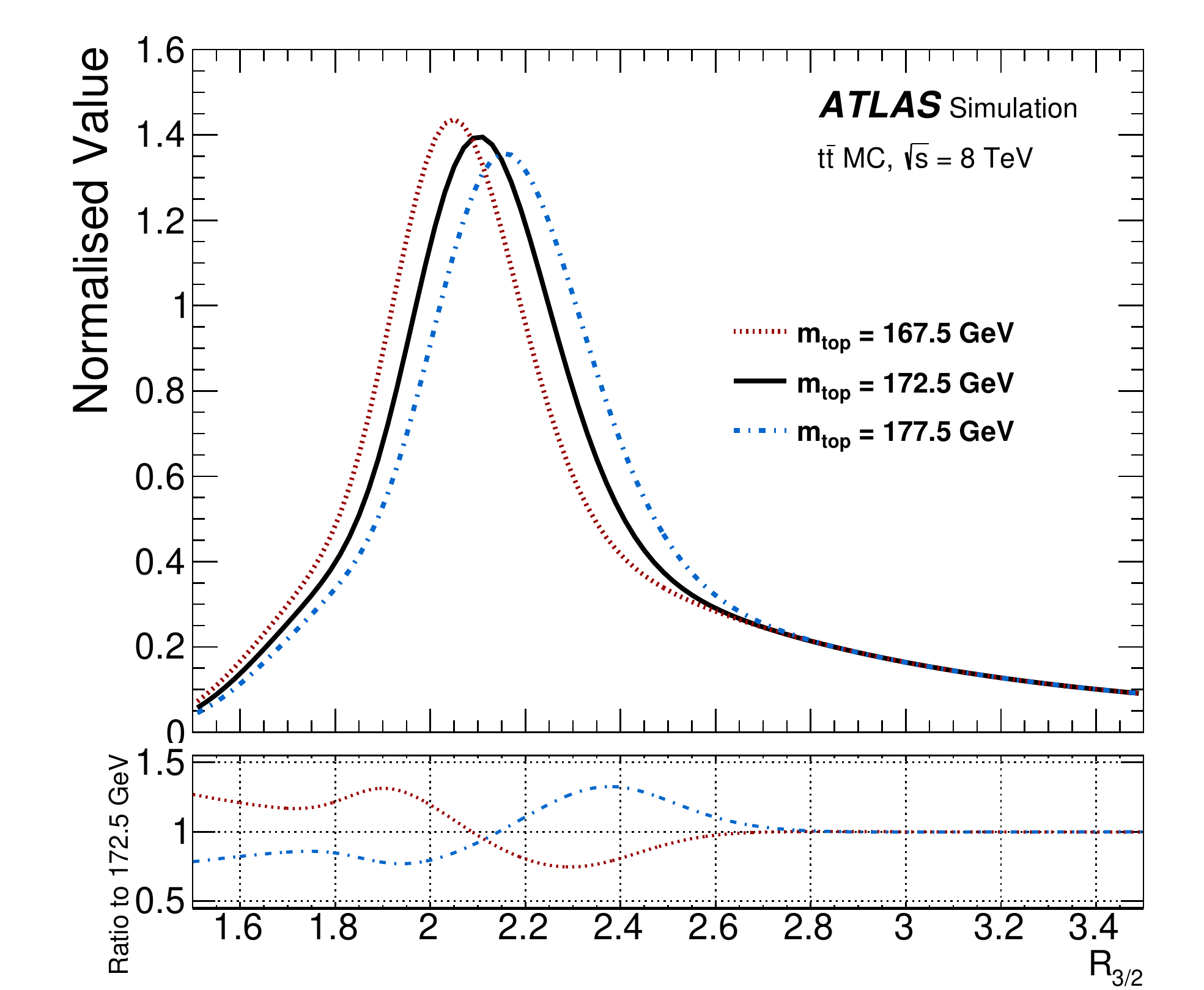}
\caption{(left) $R_{3/2}$ distribution in data compared to the sum of $t\bar{t}$ simulation and the data-driven multi-jet background. (right) Sensitivity of the $R_{3/2}$ distribution to the value of the top quark mass. The figures are taken from Ref.~\cite{Aaboud:2017mae}.}
\label{fig:ATLAShadmass}
\end{figure}

The $t\bar{t}$ final state is reconstructed using a $\chi^2$ procedure. In order to minimise the impact of the jet energy scale uncertainty, the key observable used in the measurement is the ratio $R_{3/2}\equiv m_{jjj}/m_{jj}$, where $m_{jjj}$ is the reconstructed mass of a top quark, and $m_{jj}$ is the reconstructed mass of the associated $W$-boson. The top quark mass is extracted by performing a template fit to the observed $R_{3/2}$ distribution, whose dependence on the top quark mass is indicated in Figure~\ref{fig:ATLAShadmass}. The template of the multijet QCD background is obtained in a data-driven way from a control sample. Excellent agreement is observed between data and the estimated background in the $R_{3/2}$ distribution, as shown in Figure~\ref{fig:ATLAShadmass}. The extracted value of the top quark mass is $m_{\rm Top} =$ 173.72 $\pm$ 0.55 (stat.) $\pm$ 1.01 (syst.). The dominant systematic uncertainties are found to be the jet energy scale and hadronisation modelling.

It is hoped that measurements of the top quark mass in non-standard topologies might provide a complementary source of information, ultimately resulting in reduced uncertainties when performing statistical combinations with other methods. To this end, CMS recently performed a measurement of the top quark mass in boosted topologies~\cite{Sirunyan:2017yar}.

The measurement is performed in the lepton+jets channel using the full 8 TeV dataset collected by CMS. Highly boosted, hadronically decaying top quarks are reconstructed using the Cambridge-Aachen clustering algorithm with a distance parameter $R=1.2.$ One additional small radius ($R=0.5$, reconstructed with the anti-$k_{\rm T}$ algorithm) jet from the semileptonically decaying top is also required. A veto is imposed on additional jet activity to ensure that the hadronically decaying top quark is fully merged into one jet. 

\begin{figure}[htb]
\centering
\includegraphics[height=2in]{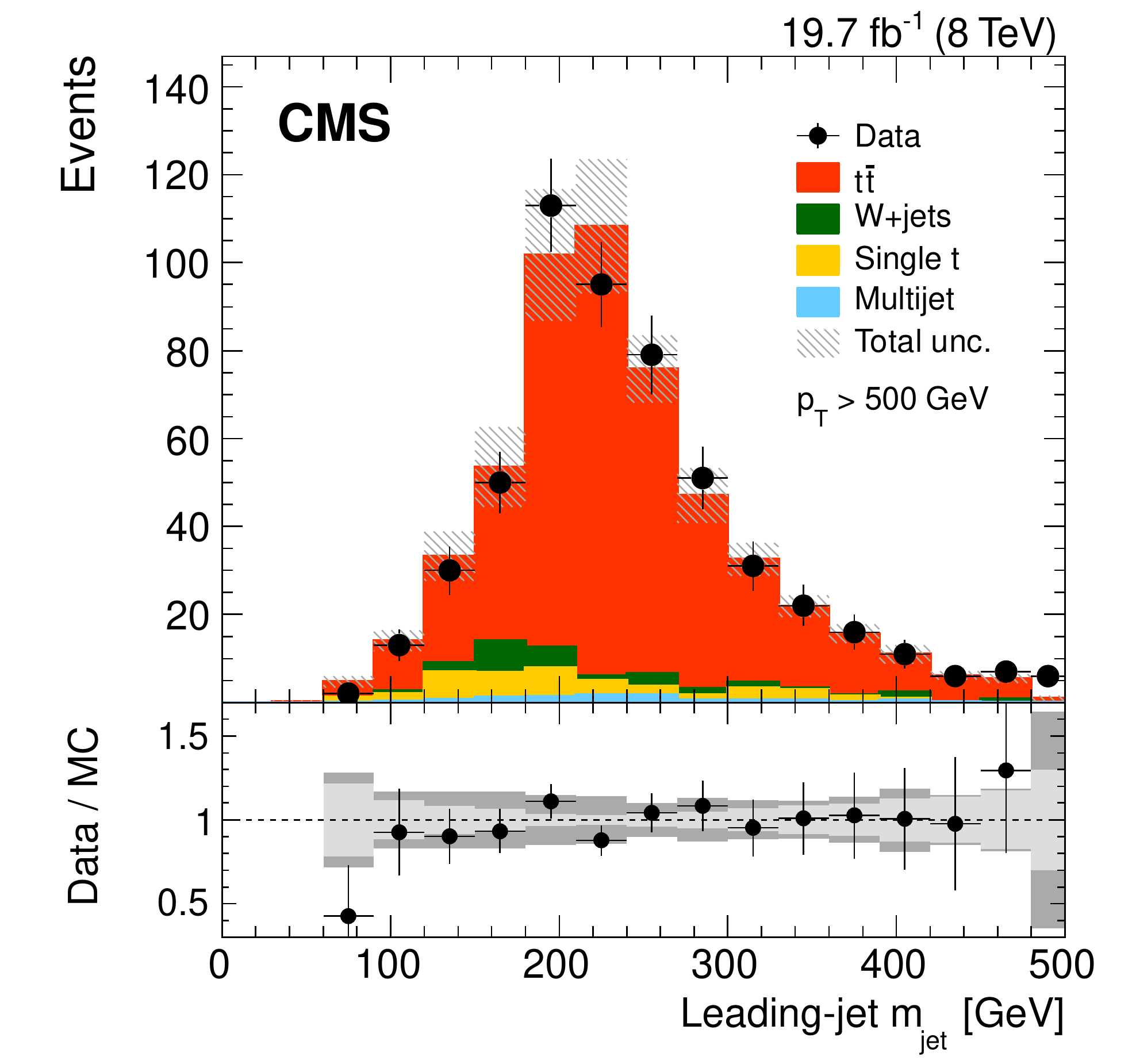}
\includegraphics[height=2in]{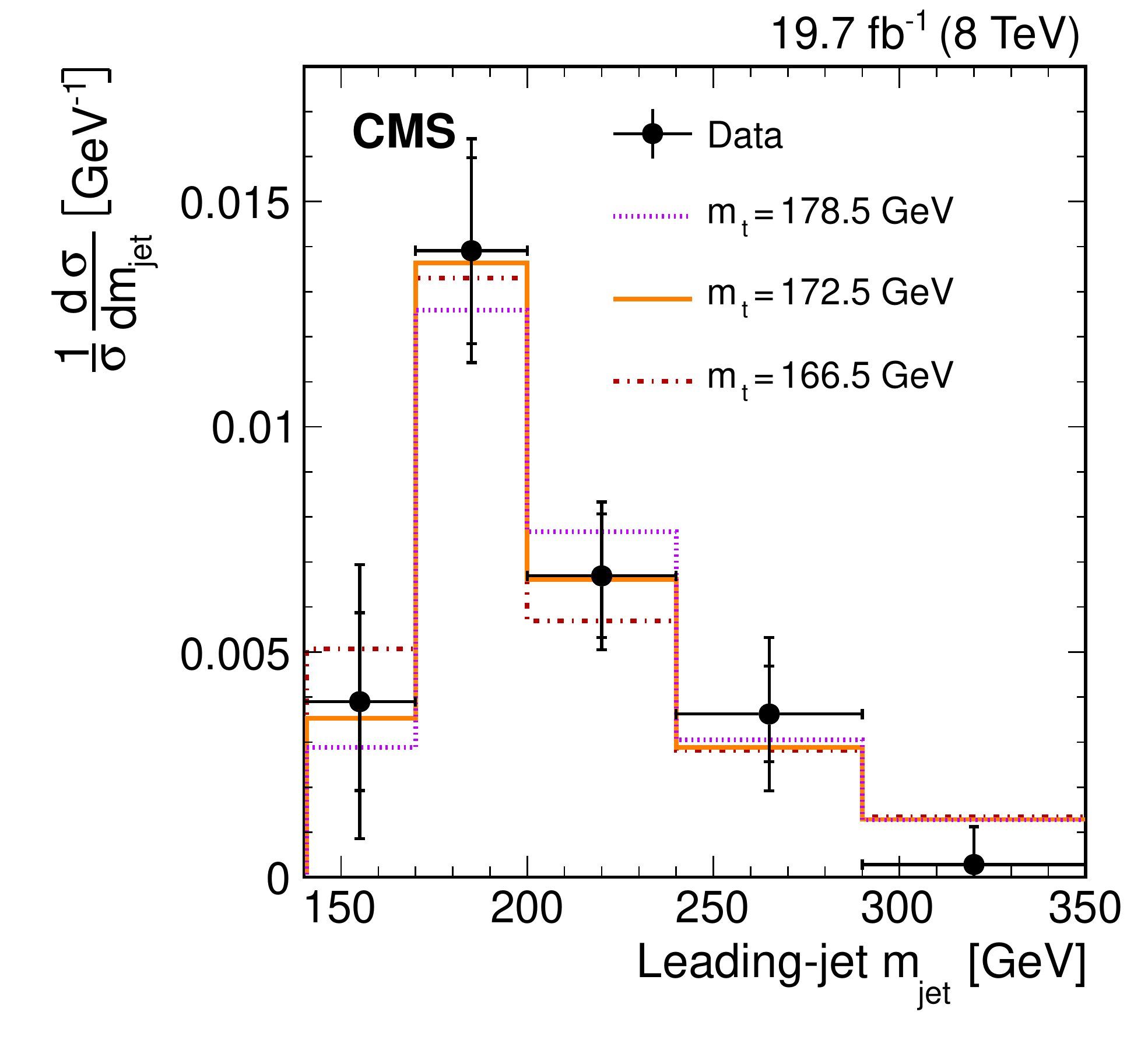}
\caption{(left) Distribution of the leading jet invariant mass in data and simulation, for jets with $p_{\rm T}>500$ GeV. (right) the normalised particle level differential $t\bar{t}$ cross section compared to predictions from Madgraph+Pythia8 samples with varying $m_t.$ The figures are taken from Ref.~\cite{Sirunyan:2017yar}.}
\label{fig:CMSboostedmass}
\end{figure}

The analysis proceeds by reconstructing the $m_{\rm jet}$ distribution of the hadronically decaying top quark, whose peak position is sensitive to $m_t.$ Figure~\ref{fig:CMSboostedmass} shows the excellent agreement between data and simulation seen for the $m_{\rm jet}$ distribution. The distribution is unfolded to particle level and the top quark mass is extracted by comparing the unfolded distribution to predictions from MadGraph+Pythia8 Monte Carlo samples with varying $m_t$, as shown in Figure~\ref{fig:CMSboostedmass}. The extracted value is $m_t = 170.8\pm 6.0$ (stat) $\pm2.8$ (syst) $\pm4.6$ (model) $\pm2.8$ (theo) GeV. The significant statistical uncertainty will decrease with the larger dataset of the $\sqrt{s}=13$ TeV run of the LHC; the primary challenge for future improvements will be reducing the $t\bar{t}$ modelling uncertainties as well as the uncertainties from the jet mass and energy scale.

In another recent publication~\cite{Sirunyan:2017idq}, the CMS collaboration measured the top quark mass in the dilepton channel, using the full 8 TeV dataset. Here the key observables used are $M^{bb}_{\rm T2}$~\cite{Barr:2010zj} and the invariant mass $M_{bl}$ of a $b$-jet and lepton arising from a semileptonically decaying top quark. Distributions of both observables are expected to exhibit endpoints whose positions depend on the value of $m_t.$ Since both variables only use $b$-tagged jets, the measurement is expected to be sensitive only to the overall jet energy scale, unlike conventional methods in the lepton+jets channel that are sensitive to flavour dependent uncertainties. In addition, since the number of jets used in the two observables is different, the observables exhibit a different dependence on the jet energy scale. Therefore, fitting them at the same time makes possible the simultaneous extraction of the top quark mass and the overall jet energy scale factor (JSF). 

\begin{figure}[htb]
\centering
\includegraphics[height=2in]{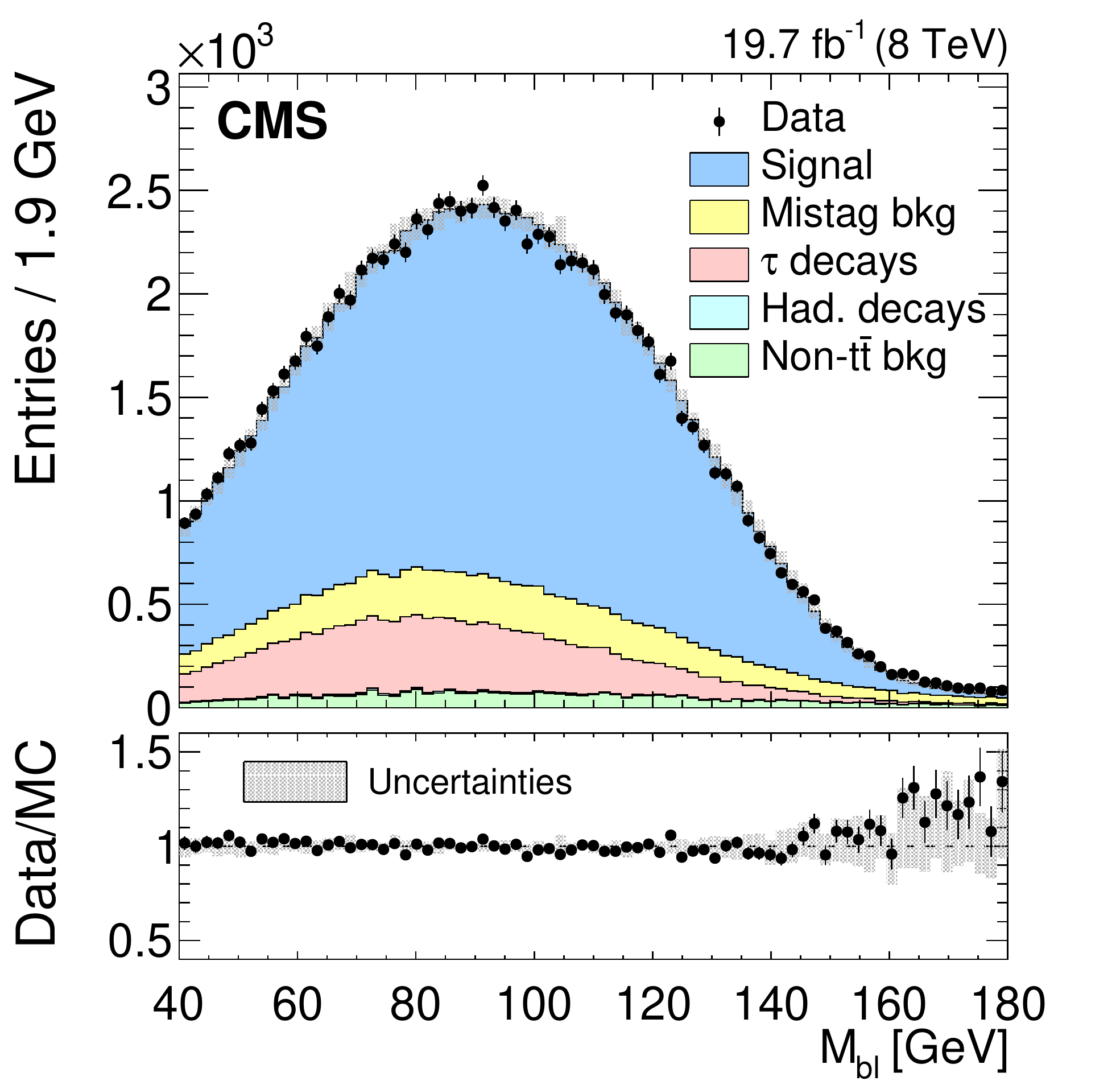}
\includegraphics[height=2in]{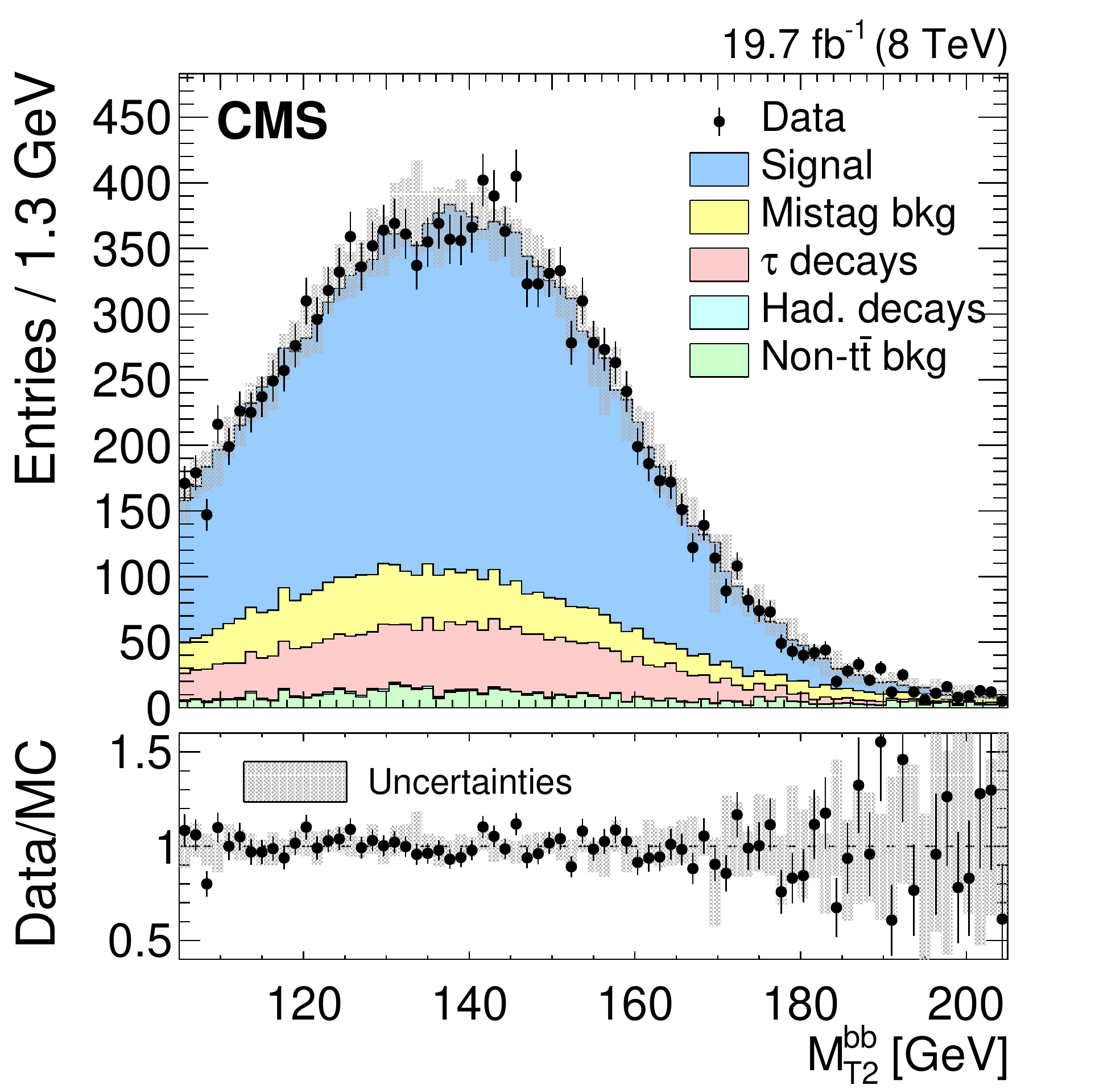}
\includegraphics[height=2in]{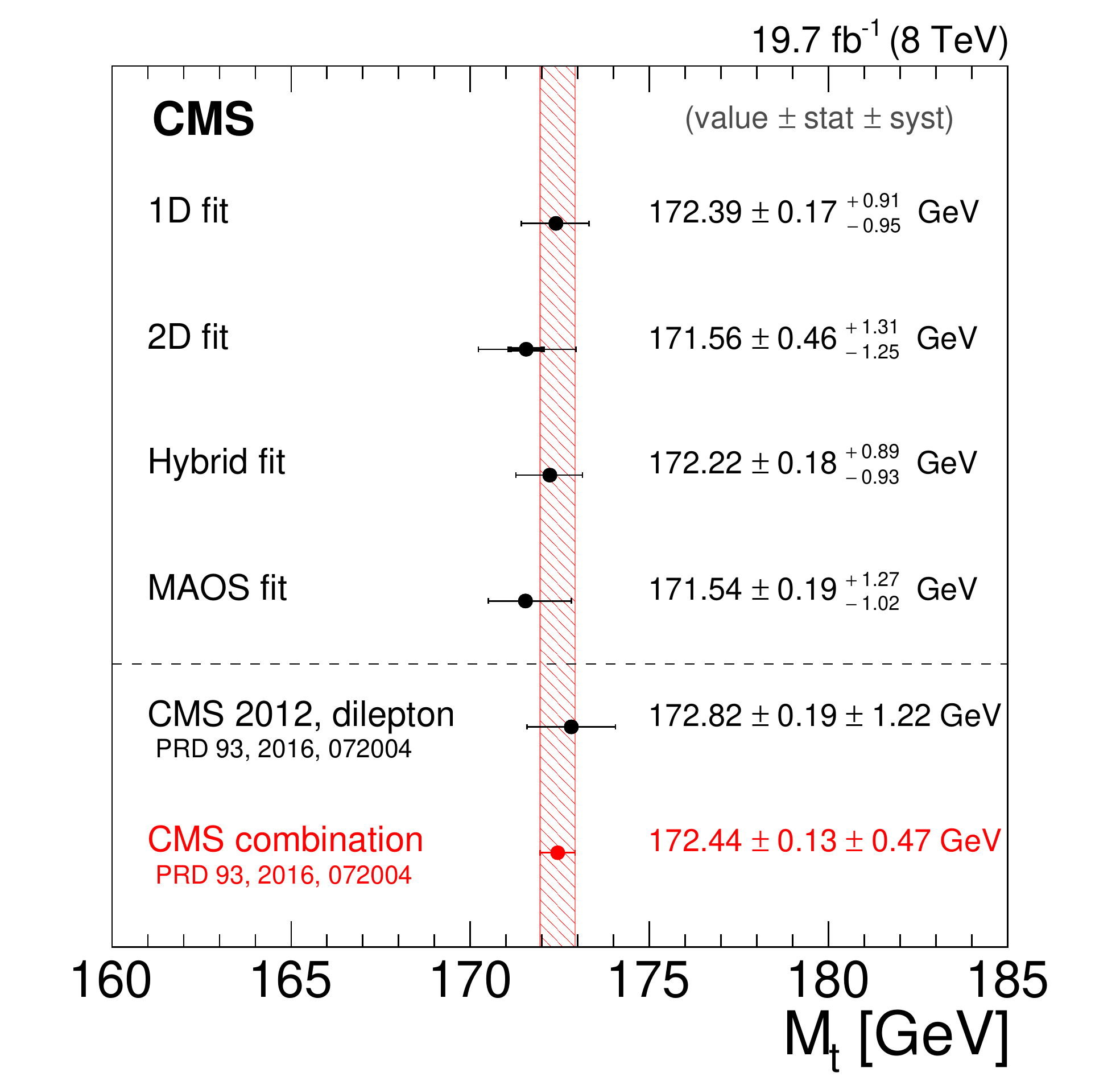}
\caption{(left) The $M_{bl}$ distribution in data and simulation (with $m^{MC}_t = 172.5$ GeV); (centre) the $M^{bb}_{\rm T2}$ distribution in data and simulation; (right) the top quark mass obtained using different fitting strategies. The figures are taken from Ref.~\cite{Sirunyan:2017idq}.}
\label{fig:CMSdilepmass}
\end{figure}

The top quark mass is extracted using an unbinned maximum likelihood fit, together with the JSF. A Gaussian process regression technique is used to obtain smooth, non-parametric distribution shapes of the observables as functions of $m_t$ and the JSF.

The good agreement between data and expectation in the $M^{bb}_{\rm T2}$ and $M_{bl}$ distributions is shown in Figure~\ref{fig:CMSdilepmass}. Various fit options are explored, including the use of a one-dimensional fit in which the JSF is fixed to unity, and using in place of $M_{bl}$ the $M_{bl\nu}$ observable with the neutrino is reconstructed by the $M_{T2}$-assisted on-shell (MAOS) reconstruction technique~\cite{Cho:2008tj}. The results obtained from the various methods considered  are also shown in Figure~\ref{fig:CMSdilepmass}. The most precise measurement of the top quark mass is obtained from a linear combination (``hybrid fit'') of the one-dimensional (where the top quark mass is fitted and the JSF is constrained to unity) and two-dimensional fits (where both the top quark mass and the JSF are fitted) and is found to be  $m_t = 172.22\pm$ 0.18 (stat) ${\ }^{+0.89}_{-0.93}$ (syst) GeV. The total uncertainty is dominated by uncertainties in modelling, the jet energy scale and $b$-quark fragmentation.

All the measurements discussed above use $t\bar{t}$ events. In Ref.~\cite{Sirunyan:2017huu}, CMS reported the first measurement of the top quark mass using single top events. This is well motivated by the fact that $t\bar{t}$ modelling uncertainties tend to form a large fraction of the total uncertainty on measured $m_t$ in $t\bar{t}$ events. By measuring the mass in single top quark events, where the modelling uncertainties are expected to be partly uncorrelated with the ones in $t\bar{t}$ events, it is hoped that a more precise measurement can be obtained in a statistical combination of the two approaches.

The measurement uses $t$-channel single top events in the 8 TeV dataset collected by CMS. The analysis is performed in the $\mu$+jets channel, selecting events with one $b$-tagged jet and one forward ($|\eta|>2.5$) jet, characteristic of $t$-channel single top production. The top quark candidate is reconstructed from the $b$-tagged jet, the muon candidate and the neutrino, whose momentum is obtained by using the missing transverse momentum and the $W$ mass constraint. 

\begin{figure}[htb]
\centering
\includegraphics[height=2in]{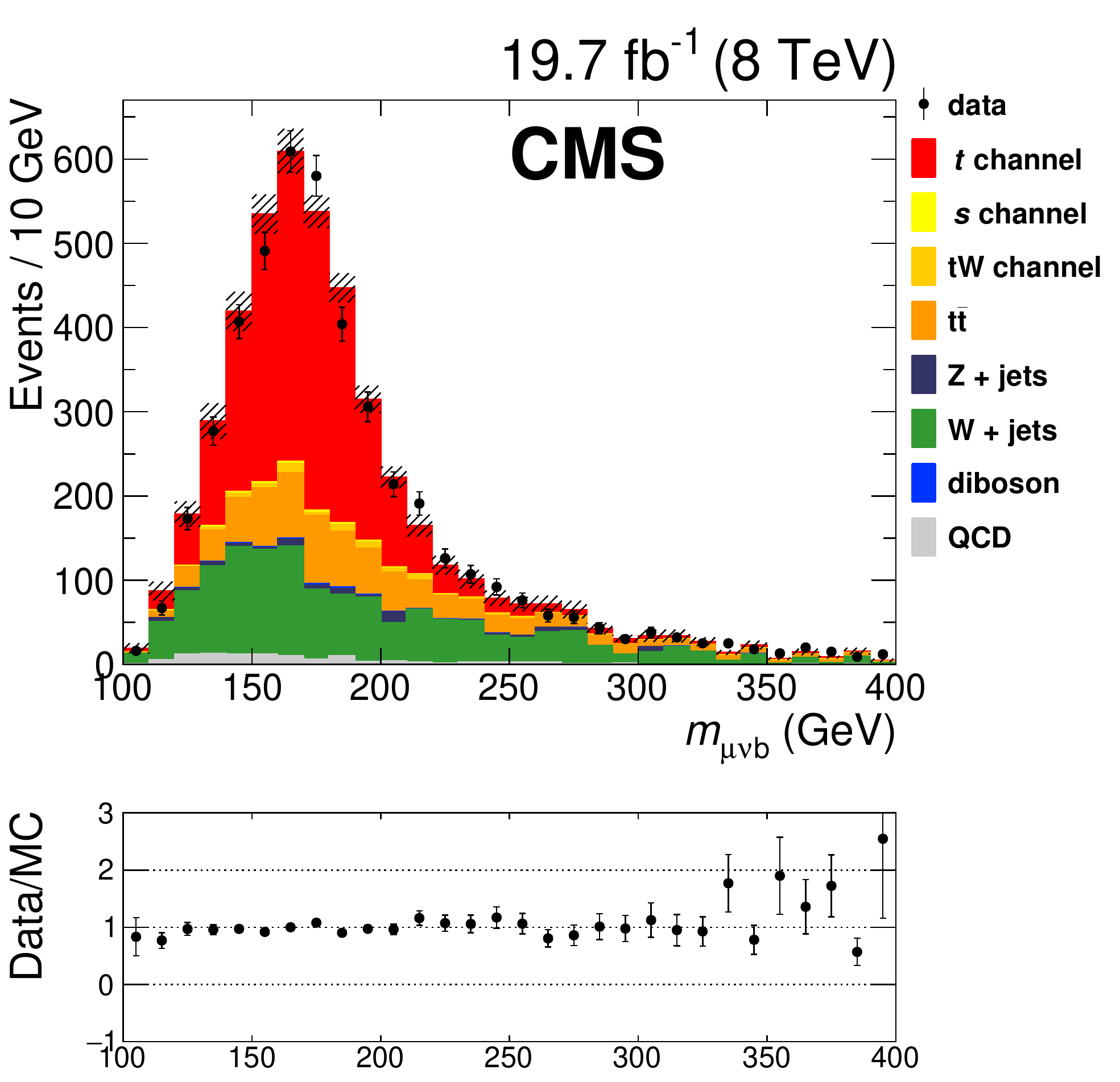}
\includegraphics[height=2in]{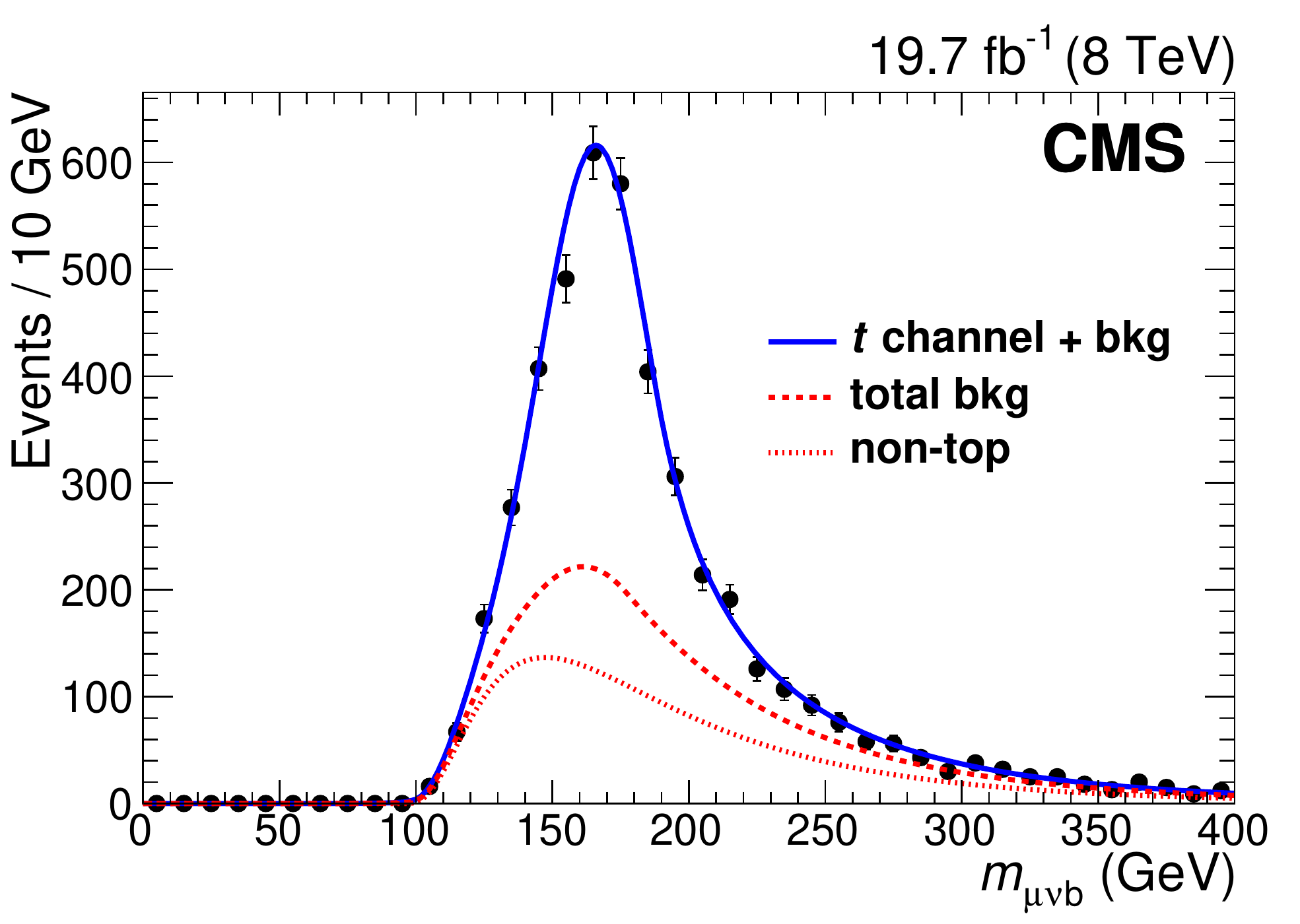}
\caption{(left) The reconstructed $m_{\mu\nu b}$ distribution in data and simulated Monte Carlo events. (right) Result of the fit to data, with the top quark and non-top-quark components shown separately. The figures are taken from Ref.~\cite{Sirunyan:2017huu}.}
\label{fig:CMSsingletopmass}
\end{figure}

The top quark mass is extracted using an extended unbinned maximum likelihood fit to the $m_{\mu\nu b}$ distribution. The shapes of the $m_{\mu\nu b}$ distribution for signal and background events is extracted from Monte Carlo simulation and modelled by appropriate functions: Crystal Ball for processes containing top quarks and a Novosibirsk function for all other processes. Figure~\ref{fig:CMSsingletopmass} shows the reconstructed $m_{\mu\nu b}$ distribution in data and simulated background and signal events, as well as the result of the fit. The value $m_t = 172.95\pm$ 0.77 (stat) ${\ }^{+0.97}_{-0.93}$ (syst) GeV is obtained from the fit. The largest uncertainties are found to be the jet energy scale and background modelling uncertainties.

While all measurements reviewed here use the 8 TeV ATLAS and CMS datasets, CMS recently also reported the first preliminary result~\cite{CMS-PAS-TOP-16-022} using data collected in 2015 at the centre of mass energy of $\sqrt{s}=13$ TeV. This analysis is performed in the muon+jets channel and uses the same method as that of the currently most precise individual measurement of the top quark mass ($m_t = 172.35\pm 0.51$ GeV, reported in Ref.~\cite{Khachatryan:2015hba}). The measured value of the top quark mass, $172.62\pm 0.38$ (stat.+JSF)$\pm$ 0.70 (syst.) GeV, is consistent with the 8 TeV result, and is still in some tension with the most precise result from the Tevatron, $m_t = 174.98\pm 0.76$ GeV.

%%%%%%%%%%%%%%%%%%%%%%%%%%%%%%%%%%%%%%%%%%%%%%%%%%%%%%%%%%%%%%%%%%%%%%%%%
%%
%%   use this format to include an .eps figure into your paper
%%
%\begin{figure}[htb]
%\centering
%\includegraphics[height=2in]{head_lhcp2017.jpg}
%\caption{ Place the caption here}
%\label{fig:figure1}
%\end{figure}
%%%%%%%%%%%%%%%%%%%%%%%%%%%%%%%%%%%%%%%%%%%%%%%%%%%%%%%%%%%%%%%%%%%%%%%%%%%

%See Figure \ref{fig:figure1} and Table \ref{tab:table1}. 

%%%%%%%%%%%%%%%%%%%%%%%%%%%%%%%%%%%%%%%%%%%%%%%%%%%%%%%%%%%%%%%%%%%%%%%%%
%%
%%   use this format to include a LaTeX table  into your paper
%%
%\begin{table}[t]
%\begin{center}
%\begin{tabular}{l|ccc}  
%Patient &  Initial level($\mu$g/cc) &  w. Magnet &  
%w. Magnet and Sound \\ \hline
% Guglielmo B.  &   0.12     &     0.10      &     0.001  \\
% Ferrando di N. &  0.15     &     0.11      &  $< 0.0005$ \\ \hline
%\end{tabular}
%\caption{ place the caption here }
%\label{tab:table1}
%\end{center}
%\end{table}
%%%%%%%%%%%%%%%%%%%%%%%%%%%%%%%%%%%%%%%%%%%%%%%%%%%%%%%%%%%%%%%%%%%%%%%%%%%

%\section{Interpretations}

%......

\section{Conclusions}
The last year has seen major progress in measurements of the properties of the top quark at the LHC experiments. The top quark polarisation and spin correlations have been measured to percent-level accuracy. The polarisations of $W$ bosons from top quark decays have also been measured with a similar precision, and strong constraints have been imposed on possible modifications of the $Wtb$ vertex. Signals of CP violation have been sought for in top quark production and decay and also in $b$-hadrons arising from top quark decays. The top quark mass has also recently been measured in a variety of final states and topologies, including all-hadronic $t\bar{t}$ decays, boosted top quark topologies and $t-$channel single top production. 

Many interesting developments are expected in the near future. Some of the measurements reviewed here - the search for CP violation in $t\bar{t}$ events and the mass measurement in boosted jet topologies are statistically limited and will benefit from the large integrated luminosity of the 13 TeV dataset. An update of the ATLAS measurement in the lepton+jets channel using the full 8 TeV dataset is expected. Combinations of the top quark mass measurements using ``traditional'' methods with measurements in novel topologies will further reduce the uncertainty in $m_t.$  In the longer term, an important challenge will be reducing the impact of modelling uncertainties using new Monte Carlo techniques and experimental strategies.

%...... 

%%  if necessary
\Acknowledgements
I am grateful to Frederic Deliot, Reinhard Schwienhorst and Rebeca Gonzalez Suarez for fruitful discussions.

\end{document}

%% file: econfmacros.tex
\def\beq{\begin{equation}}
\def\eeq#1{\label{#1}\end{equation}}
\def\eeqn{\end{equation}}

\def\beqa{\begin{eqnarray}}
\def\eeqa#1{\label{#1}\end{eqnarray}}
\def\eeqan{\end{eqnarray}}

\let\bar=\overbar

\def\Dslash{\not{\hbox{\kern-4pt $D$}}}
\def\dslash{\not{\hbox{\kern-2pt $\del$}}}

\def\msb{{\bar{\ssstyle M \kern -1pt S}}}